\documentclass[]{emulateapj}
\usepackage{graphicx}
\newcommand{\msun}{$M_{\sun}$}
\newcommand{\rsun}{$R_{\sun}$}
\newcommand{\vsini}{$v\sin i$}
\newcommand{\kms}{${\rm km}\ {\rm s}^{-1}$}
\def\ga{\mathrel{\hbox{\rlap{\hbox{\lower4pt\hbox{$\sim$}}}\hbox{$>$}}}}
\shorttitle{Planet Accretion Creating Red Giant Rapid Rotators}
\shortauthors{Carlberg, Majewski, \& Arras}

\begin{document}
\title{The Role of Planet Accretion in Creating the\\ Next Generation of Red Giant Rapid Rotators}
\author{Joleen K. Carlberg\altaffilmark{1}, Steven R. Majewski\altaffilmark{1},  and Phil Arras\altaffilmark{1}}
\altaffiltext{1}{Department of Astronomy,
University of Virginia,
Charlottesville, VA 22904, USA
Email: jkm9n@virginia.edu, srm4n@virginia.edu, pla7y@virginia.edu}

\begin{abstract} 
Rapid rotation in field red giant stars is a relatively rare but well-studied phenomenon; here we investigate the potential role of planet accretion in spinning up these stars. 
Using Zahn's theory of tidal friction and stellar evolution models, we compute the decay of a planet's orbit into its evolving host star and the resulting transfer of angular momentum into the stellar convective envelope. This experiment assesses the frequency of planet ingestion and rapid rotation on the red giant branch (RGB) for a sample of 99 known exoplanet host stars.
We find that the known exoplanets are indeed capable of creating rapid rotators; however, the expected fraction due to  planet ingestion is only $\sim$10\% of the total seen in surveys of present-day red giants.
Of the planets  ingested,  we find that those with {\it smaller} initial semimajor axes are more likely to create 
rapid rotators because these planets are accreted when the stellar moment of inertia is smallest.   We also  find that many planets may be ingested prior to the RGB phase, contrary to the expectation that accretion would generally occur when the stellar radii expand significantly as giants.  Finally, our models suggest that the rapid rotation signal from ingested planets is most likely to be seen on the lower RGB, which is also where alternative mechanisms for spin-up, e.g.,  angular momentum dredged up from the stellar core, do not operate.  Thus, rapid rotators on the lower RGB are the best candidates to search for
definitive evidence of systems that have experienced planet accretion.
\end{abstract}

\section{INTRODUCTION}
The rotation rates of solitary field giant stars as a function of temperature show a sharp transition from fast rotation to slow rotation redward of the G0III spectral type  \citep{Gray89}.  The majority of stars redward of  the transition are characterized by a mean rotational velocity of  $\sim$~2~\kms\ \citep{deMed96a}.
However, a small percentage of these red giant stars are found to have rapid rotation, which is generally defined as \vsini\ $\ge$ 10~\kms.  Such high rotation rates are unexpected with normal  
stellar evolution because  angular momentum is conserved as the stars expand.
Moreover, \cite{Gray89} argues that the sharp transition from high to low rotation at G0III indicates an ancillary braking mechanism  caused by a magnetic dynamo effect created as 
the stellar convection zone deepens, further slowing the rotation; this mechanism renders rapid rotation even less likely in red giants.

With the lack of obvious progenitors under normal evolutionary processes, these unusual, rapidly rotating red giants must be {\it spun up} by some mechanism, either
internal or external.  One proposed internal mechanism is angular momentum dredge-up \citep{simon89,fekel93}, which relies on the star having  a rapidly spinning  core that rotates independent of the envelope.  The core acts as an angular momentum reservoir that is tapped  as the convective envelope extends  deep into the star during first dredge-up  on the  red giant branch (RGB).

An alternative mechanism is a gain in angular momentum from an external source, which, in the absence of a cluster environment or stellar companion, would be
the orbital angular momentum delivered by a brown dwarf or planet. 
\cite{pete83} were the first to consider planets as a source of angular momentum in evolved stars.    Later studies \citep{soker98,soker00} invoked planet-induced mass loss as the possible ``second parameter''  determining horizontal branch (HB) morphologies. 
In studies of red giant chemistry, \cite{alexander67} first invoked planet engulfment by red giants to explain lithium enhancements in some of these stars.  Since these early studies,  the idea of planet engulfment has been invoked repeatedly explain  both lithium  enhancement and rapid rotation observed in RGB stars \citep[][and references therein]{siess99,reddy02,drake02,carney03,denissenkov04}.
\cite{siess99} modeled the accretion of sub-stellar companions by red giant stars and calculated observational
signatures of the accretion in those stars, such as infrared excess and $^7$Li enrichment. From the actual occurrence of these predicted observational signatures found in the red giant population, they estimated that  4--8\% of Sun-like stars must host planets  if most planet-hosting (PH) stars have significant interactions with their planets while on the RGB.   
\cite{livio02}, on the other hand, used the extrasolar planets known at the time to estimate that at least 3.5\% of red giants  would have their evolution  ``significantly affected'' by planets,  with  rotation-induced mass-loss being the most significant alteration.  A comprehensive discussion of the effects planets may have on a star's evolution from the RGB through the HB and asymptotic giant branch (AGB) to the planetary nebula phase is given in \cite{soker04}.

In this study, we expand on the idea of \cite{livio02} of  using the known characteristics of planet-hosting stars to predict future interactions between the RGB stars and planets. We are particularly interested in trying to account for the frequency of RGB stars with rapid rotation by planets depositing angular momentum into their parent stars.
Although we focus on planets, we do note that brown dwarfs and  low-mass stellar companions are also capable of spinning up giants.   However,  brown dwarfs turn out to be unlikely angular momentum sources because few stars have  brown dwarfs orbiting near enough to their stars to tidally interact significantly with their host stars \citep{grether06}.
Low-mass stars are twice as common as massive planets \citep{grether06}; however, their higher mass affords them a higher survival rate in common envelope evolution.  Therefore,  they may not explain rapid rotation in samples of field giant stars that were selected to be companionless.   We return to the issue of  low-mass stars in  \S \ref{sec:other}.

Using  {\it known} properties of PH stars as a basis for exploring giant rapid rotation is advantageous for a number of reasons.  First, it is not immediately obvious how some parameters affect
the probability of a star becoming a future rapid rotator.  As an example, consider the role of the semimajor axis of the planet's orbit.  The orbital radius must be sufficiently small that the planet is eventually engulfed by the star, but within this criterion one would naively suspect that planets with the largest possible orbital radius
(and thus largest angular momentum for a given mass) would maximize the probability of spinning up the star. On the other hand, as we show here, planets orbiting at larger radii will be engulfed later in the evolution of the star, when the stellar moment of inertia is larger and subsequent spin-up is smaller.  Consequently, whether or not a PH star can
become a rapid rotator depends on both the angular momentum available in the planetary orbit as well as when in the course of the host star's evolution the planet is actually accreted.
 The relative values of these properties are specific to individual planetary systems; therefore, using the properties of {\it known} PH stars incorporates the natural distributions of both the stellar and planetary properties, all of which affect the census of rapid rotators. By combining the distributions of PH stars and planets with stellar evolution models that allow us to track the individual evolution of the PH stars into the RGB phase,  we can begin to ask questions about the relationship between PH stars and the current population of red giant rapid rotators.  How many PH stars will become rapid rotators?  What is the lifetime of a red giant rapid rotator that is created through planet accretion?  Is the number and lifetime of rapid rotators from expected planet ingestion enough to account for the number of observed rapid rotators in the current red giant population?  

These are questions we address here. We begin by describing the properties of current red giant rapid rotators and PH stars  as well as our adopted stellar evolution models in \S \ref{sec:stellprop}.  In \S \ref{sec:evolve}, we present our model for the orbital tidal evolution and ingestion of exoplanets; the results of our modeling are presented in \S \ref{sec:results}.
Finally,  in \S \ref{sec:discuss} we discuss the possible effects of observational biases introduced by different exoplanet detection methods as well as the unmodeled effects of mass loss and magnetic dynamos.

\section{STELLAR PROPERTIES}
\label{sec:stellprop}
\subsection{Frequency of Red Giant Rapid Rotators}
\label{sub:kstars}
The peculiarity of rapid rotation among the red giant population has been well studied and documented by many authors.  While the exact value of the rotational velocity separating ``rapid
rotators'' and ``normal rotators'' is somewhat arbitrary,  we follow the literature standard of  defining a rapid rotator as any red giant with \vsini~$\ge$~10~\kms. In the catalog of rotational velocities from \cite{glebocki00}, rapid rotators account for about 5\% of the red giant population.   In a sample of 900 stars, \cite{deMed96a}  find that less
than 5\% of the red giants have \vsini\ $>$~5~\kms\ and faster rotators are even rarer.   A recent study by J.~Carlberg et al.\ (in preparation)  of almost 1300 distant
K giants found that less than 6\% of their sample were rapid rotators (depending on the number of undetected binaries that  masquerade as  rapid rotators), while the study of almost 750 nearby single giants from Hipparcos, by \cite{massarotti08a}, yields only 1\% of giant stars with rapid rotation.   The general consensus, as evidenced by the four studies mentioned, is that the frequency of rapid rotation in the red giant population is  only a few percent.  
\begin{figure}
\epsscale{1.0}
\plotone{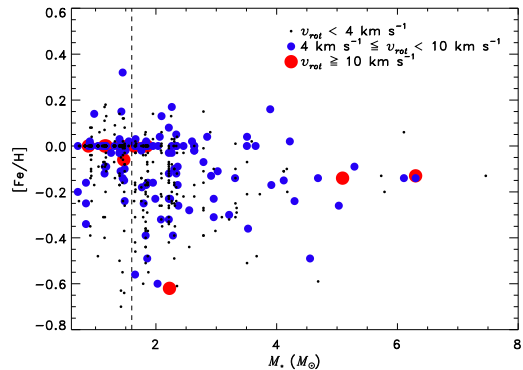}
\caption{Stellar mass and metallicity for a sample of  low-mass single red giants from \cite{massarotti08a}.  
The different sized circles indicate bins of \vsini\ $<$~4~\kms\  (small),  4 $\le$ \vsini\ $<$ 8~\kms\ (medium), and \vsini\ $\ge $ 8~\kms \ (large).
The vertical line indicates the maximum stellar mass probed by our PH sample. }
\label{fig:smassfeh}
\end{figure}

In this work, we compare predictions of the frequency of rapid rotation from planet ingestion (as well as trends of rapid rotation frequency with stellar properties)  with the sample of observed red giants in the \cite{massarotti08a} survey  because the Hipparcos distances allow
accurate estimates of absolute stellar luminosities and masses.  Figure \ref{fig:smassfeh} shows the masses and metallicities of 
669  stars   from this survey grouped into three velocity bins.  Included are all stars from the survey with masses between 0.7 and 
10~\msun, temperatures less than $\sim$5550~K, and $\chi ^2~<~0.001$ (as suggested by the authors of that paper  to avoid possible undetected binaries).   
 In this sample, ten stars (1.5\%) show rapid rotation,  122  (18.2\%) are in an intermediate \vsini\ bin (arbitrarily defined as $4~\le$~\vsini$~<~10$~\kms), and the remaining stars  are slow rotators.  
 For reference, the vertical line shows that maximum mass probed by the PH stars described in the next section.
In the stellar mass range probed by our exoplanet host star sample, 5 stars (1.3\%) show rapid rotation,  40 (10.7\%) are in the intermediate \vsini\ bin, and the remaining stars  are slow rotators.  Note in Figure \ref{fig:smassfeh} that there is a paucity of both rapid and intermediate rotators for the most metal poor stars for the  low-mass range to the left of the line.

\subsection{Exoplanet Data}
\label{sec:exodata}
The properties of exoplanet host stars were downloaded  from  {\it The Extrasolar Planets Encyclopaedia}\footnote[2]{http://www.exoplanet.eu/catalog.php, database for ``All Candidates detected,'' accessed
 on 2008 July 17.}, which is a compilation of both the known orbital parameters  of exoplanets and the physical properties of their host stars.
We selected all those exoplanet host stars for which  masses and orbital radii of their planets were available, for which stellar masses were available,
which were known to be luminosity class V stars  and thus on the main sequence (MS), and which would evolve to the RGB in less than 13.5 Gyr  (a total of 99 systems with 115 planets).  
A second sample adds all of the stars with no known luminosity class but that meet the rest of our selection criteria (38 additional planets in 31 additional systems), and for these, we simply assume that they are likely to be  MS stars.  In the following paragraphs we will give statistics  for the first sample followed by those of the second sample in parentheses.  The PH stars have [Fe/H] ranging from -0.32 to 0.50 dex (-0.68 to 0.50) with an average of 0.09 dex (0.10), while the masses range from 0.91 to 1.43~\msun\ (0.87 to 3.9) with an average of 1.1~\msun\ (1.2).

Eccentricities of the planetary orbits were used in our calculations when available, and we assumed circular orbits in the absence of such a measurement.
Because the majority of exoplanets have been discovered by the radial velocity method, the observed planetary masses suffer from an inclination ambiguity.   However, observable stellar rotational
velocities suffer from the same ambiguity.  Therefore, in using the minimum planetary mass ($M_{p}\sin i$)   we can compute the {\it observable} rotational velocity, \vsini,  for the PH stars that ingest planets to directly compare to \vsini\ measurements of actual red giant stars.  This substitution is valid under the assumption that the planets' orbital axes are aligned with their stars' rotational axes.  For comparison, this assumption is valid in our own solar system because all of the planets' orbital axes are
tipped by only  few degrees with respect to the ecliptic, and the ecliptic itself is tipped by only 7$^{\circ}$ to the Sun's rotational axis.

 \subsection{Evolution Models}
 \label{sec:evmod}
The evolution of a star is mostly determined by  its mass and composition alone \footnote[3]{Rotation also plays an important role; however, grids of evolution models explicitly accounting for rotational effects are not currently available for the range of stellar masses of the stars in our study.  Some unmodeled effects of rotation are addressed in \S \ref{sec:mdot}}. 
 All of the PH stars in our sample have mass measurements, and all except 14 
 (18) have measured metallicities. We assume solar metallicity for those without such a measurement.  With these two properties we assign each star a stellar evolution track.  We use the stellar evolution models of \cite{giard00}, which provides a grid of models with metallicities ranging from $Z=0.0004$ to $Z=0.03$ ([Fe/H]~$\approx$~-1.74 to 0.18 dex) and a stellar 
mass range  of 0.15--7.0~\msun.  The models give values of the stellar age, effective temperature and luminosity (hence radius),  evolution stage, and $q_{disc}$ as a function of time. 
\citep[See][for a complete description of the models and all available global parameters.]{giard00}

The parameter  $q_{disc}$  is the mass fraction contained within the radius where the chemical composition first differs from that at the surface.  We use it as an indicator of the amount of mass in the homogenized
convective envelope,  given by $M_{env}=M_*(1-q_{disc})$.   The mass within the convective envelope is important because this is the region of the star  where tidal dissipation occurs \citep{zahn77};
it affects the rate at which a planet spirals in and is the region where the angular momentum is deposited.
The  $q_{disc}$ parameter works well as an indicator of mass in the convective envelope from the MS turn-off up to first dredge-up on the RGB because  the convection zone  is responsible for keeping the star homogenized from the surface to the discontinuity, $q_{disc}$.  However, once first dredge-up ends the base of the convection zone recedes back towards the surface, while the star remains chemically homogenized to the deepest point previously reached by the convection zone.  At this evolutionary stage, $q_{disc}$  no
longer traces the convection zone.  It remains static until the outward-moving hydrogen burning shell reaches this point, after which $q_{disc}$ follows the H-burning shell.
Thus,  we find that $q_{disc}$ becomes an imperfect tracer of the convective envelope after first dredge-up  and at that point gives an {\it overestimate} of the mass in
the convective envelope. This leads to an {\it underestimate} of the stellar \vsini\ after first dredge-up for stars that accrete planets.
We also find that $q_{disc}$ is an imperfect tracer of the convective envelope on the main sequence, where the mass in the convective envelope is generally on the order of a few percent of the stellar mass while  $q_{disc}$ would give a value between 20-90\%.  For each stellar mass, we use a constant value for the fraction of mass in the convective envelope when the star is on the main sequence, which we take from the flat regions  of the plots of stellar envelope mass versus age in Figure 1 of  \cite{murray01}. 

\section{EVOLUTION OF STELLAR ROTATIONAL VELOCITY} 
\label{sec:evolve}
\subsection{Angular Momentum Budget in Planetary Orbits}
The observed rotational velocity of the PH stars will evolve as the moment of inertia of the stellar envelope evolves or if the total angular momentum in the convective envelope changes.  The moment of inertia in a stellar convective envelope, assuming  an $n=3/2$ polytrope  model,
is given by $I_{env}=(1/8)M_{env}R_{*}^2$ , where $M_{env}$ is the envelope mass and $R_{*}$ is the stellar radius. The moment of inertia will steadily increase as the star ascends the RGB, primarily from the increase in $R_{*}$.  
For simplicity, we assume that the total angular momentum in the envelope can only change if a planet is accreted; we ignore mass lost during the stellar evolution, which is also ignored in the stellar evolution models.  When a planet is accreted, the change in the stellar rotational velocity, $\Delta v_{rot}$,   from the deposition of the planet's orbital angular momentum can be calculated as
\begin{equation}
\Delta v_{rot}=\frac{8[ M_{p} \sqrt{GM_{*}a_{p}(1-e^{2})} ]} {M_{env}R_{*}}.
\label{eqvrot}
\end{equation}
The terms and factors  in Equation (\ref{eqvrot}) include the planetary mass ($M_{p}$), semimajor axis ($a_{p}$), and  eccentricity ($e$), the stellar mass ($M_{*}$),  convective envelope mass ($M_{env}$), and radius ($R_{*}$), and the gravitational constant ($G$).  The expression in square brackets  in Equation (\ref{eqvrot}) is the initial orbital angular momentum of the planet.

\begin{figure}
\epsscale{1.0}
\plotone{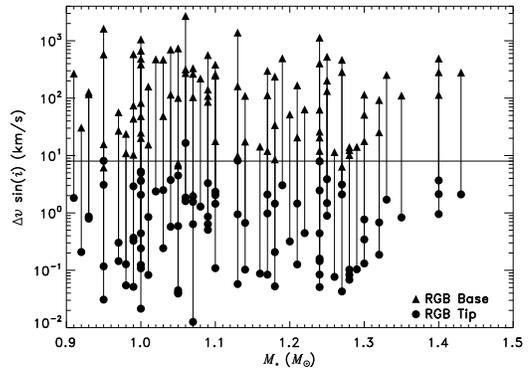}
\caption{Additional rotational velocity that PH stars would gain on the red giant branch if  the orbital angular momentum of all their respective planets with $a_p <$ 5 AU is deposited into the stellar convective envelopes.  This is calculated both when the star is at the base of the RGB (triangles) and when it is at the tip of the RGB (circles).  Solid vertical lines connect  the results for the two evolution stages for the same star.  Stars gaining more than 8 \kms\ of added rotational velocity, indicated by the horizontal line, would be considered rapid rotators. In cases where connecting
lines overlap (e.g., the three stars at 1.4 \msun), upper triangles connect to upper circles and lower triangles to lower circles.}
\label{fig:vrotpot}
\end{figure}

 Planets sufficiently close to the star will experience a strong hydrodynamical drag force in the red giant's envelope, leading to orbital decay.  Tidal friction with the red giant's envelope causes orbital decay for larger orbits where hydrodynamical effects are negligible.  All planets with semimajor axes less than  a radius, $r_{in}$, will be accreted.  \cite{soker98}  reports  that a good approximation for determining which planets  will spiral in during a star's ascent up the RGB is $r_{in} \sim5 R_{tip}$, where  $R_{tip}$ is the stellar radius at the tip of the red giant branch.  This conclusion was based on tidal equations originally presented in \cite{Soker96} and is valid
for stars with masses between  0.8 and 2.2~\msun.    For a solar mass star that reaches 200 $R_{\sun}$ at the RGB tip, all planets within $\sim$5 AU constitute an angular momentum
reservoir that will be accreted during the RGB lifetime based on Soker's approximation.  Applying this approximation to the PH stars, we can calculate the angular momentum and $\Delta v_{rot}$ 
gained from accreted planets with $a_p < 5$~AU. In Figure \ref{fig:vrotpot}, we plot this  $\Delta v_{rot}$, calculated when the PH star is both at the base of the RGB  and at the tip of the RGB.  Points for these two evolutionary stages for the same PH star are connected by a vertical line.  On average,  an additional 8~\kms\ in rotational velocity, indicated with the horizontal line, would put planet-ingesting giant stars (which generally rotate at $\sim$2~\kms\ before ingestion)  into the rapid rotator regime. This analysis illustrates two important facts.  First, the angular momentum in planetary orbits is sufficient to create RGB rapid rotators in most cases.  Second, the time of planet accretion with respect to the stellar evolutionary phase is critical because the same planet  that can create a rapid rotator if it is accreted at or near the RGB base will not 
create a rapid rotator if it is accreted when the star is at the RGB tip.  Additionally, a star that does become a rapid rotator will ``spin down'' during its ascent up the RGB.  Consequently, the point at which the planet is engulfed and the time that a star spends as a rapid rotator are very relevant to understanding how planet accretion contributes to the rapid rotator population. 

\subsection{Orbital Decay by Tidal Friction}
To find when the PH stars will actually accrete their planets, we followed the tidal evolution of the individual planetary orbital radii through the evolution of the PH star from the zero-age main sequence (ZAMS) to the RGB tip. Tidal interactions
can cause a planet to spiral into the star if the rotation frequency of the star is less than the orbital frequency of the planet.  (For the Sun, this condition would hold for planets orbiting within $\sim $ 0.2 AU, whereas
for a 15 $R_{\sun}$ red giant  with \vsini~=~2~\kms, this is true for planets out to 1 AU.)
 In such a scenario, tides raised on the star by the planet will lag behind the line connecting the centers of the two bodies, and the star will experience a tidal torque that will act in the direction that would increase the rotation frequency of the star.  This spin-up drains angular momentum
from the planet's orbit, and the planet consequently moves closer to the star. The amount of lag and the strength of the torque  depend on the rate that energy is dissipated in the tidally distorted body, and we adopt the model for turbulent dissipation presented in \cite{zahn77}. The time-scale for the inward spiral of the planet depends sensitively on the separation of the star and planet. We integrate the tidal orbital decay equations by  \cite{vp95} and \cite{zahn77}  to find
the expression for the maximum orbital separation, $r_{in}$, of accreted planets, which can be computed for every time step in the stellar evolution models. This separation is given by   
\begin{equation}
\label{eq:a}
r_{in}/R_{\sun}=\left( \frac{54.4}{9}(1+23 e^2) \left(\frac{M_*}{M_{\sun}}\right)^{-1}\left(\frac{ M_p}{M_*}\right)  I(t)\right)^{1/8}. 
\end{equation}
 The term $I(t)$ is the integral of time-dependent stellar properties that affect the tidal evolution, calculated in years from the ZAMS ($t' = 0$) to the evolutionary time step at $t$; it is described by Equation (5) of \cite{vp95} and reproduced here:
\begin{equation}
\label{eq:ioft}
I(t)= \int_0^t  \left( \frac{T_{eff}(t')}{4300 K} \right) ^ {4/3}\left( \frac{M_{env}(t')}{M_{\sun}}\right)^{2/3} \left(\frac{R(t')}{R_{\sun}} \right)^8dt'
\end{equation}
We compute the value of $I(t)$ for every time step of the stellar evolution models
through numerical integration.

\begin{figure}
\epsscale{1.0}
\plotone{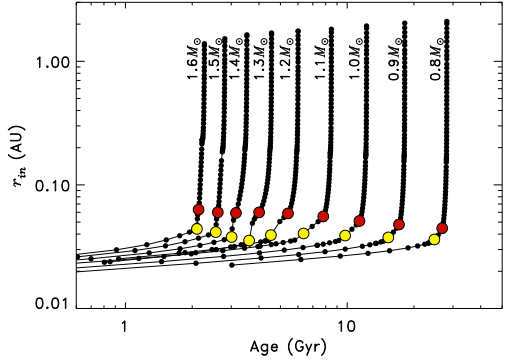}
\caption{Accretion radius, $r_{in}$,  calculated using Equation (\ref{eq:a}) for a Jupiter mass planet in a circular orbit around an evolving host star as a function of the stellar age.  Solar metallicity stellar evolution tracks  for stellar  masses ranging between 0.8 \msun\ and 1.6 \msun\ were used to follow the evolution from ZAMS to the RGB tip.  Solid lines connect the individual time steps for each mass model, and evolution progresses from bottom left to upper right. The two gray points on each track denote special evolution stages.  The yellow point  is the exhaustion of central hydrogen  and the  red point  is the base of the RGB.}
\label{fig:ain}
\end{figure}

\begin{figure}
\epsscale{1.0}
\plotone{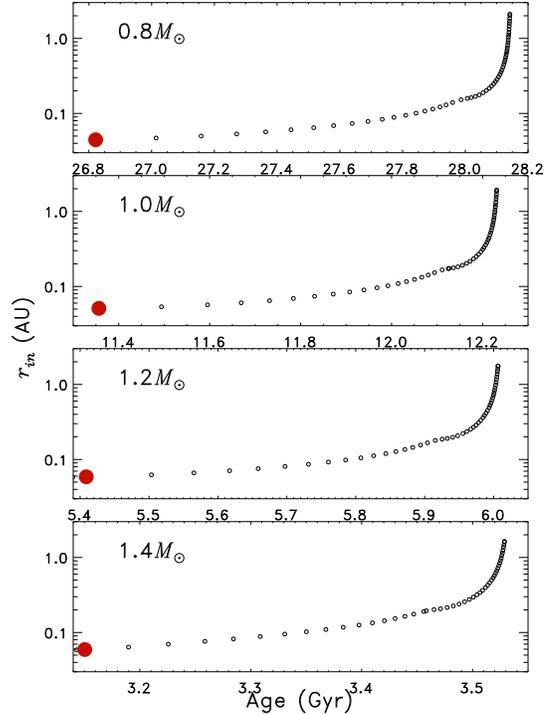}
\caption{ Expansion of the red giant branch evolution  for four of the models presented in the previous figure.  The 0.8, 1.0, 1.2, and 1.4 \msun\  tracks are shown.  The large point indicates the base of the RGB.  Note the different time-scales of each panel.}
\label{fig:ainzoom}
\end{figure}

The exoplanets are accreted by the host star in our models when $a_p= r_{in}$.   
Because the tidal decay timescales  in \cite{zahn77} depend  on  $(a_p/R_*)^8$, the planet spends very little time in the region between its original separation and the stellar surface; thus,
the angular momentum is deposited almost instantaneously compared to the evolutionary timescale of the star.
The planets will deposit 40-60\% of their angular momentum simply by moving from their initial separation to $R_*$; we assume the remaining angular momentum is added as the planet quickly evaporates in the envelope  (see \S \ref{sec:evap} for a discussion of this assumption). 
In this experiment,  the evolution is followed up to the first ascent on the red giant branch.  All of the PH stars are of low enough mass that they go through the helium flash prior to landing on the HB. The helium flash is poorly understood and is not treated in evolution models so we limit our analysis to the better understood parts of stellar evolution.  In Figure \ref{fig:ain}, we illustrate how $r_{in}$ increases during the stellar evolution for solar metallicity stars with a Jupiter-mass planet in a circular orbit. We show the evolution from ZAMS to the tip of the RGB for stars with masses between 0.8 and 1.6 \msun.  For stars of the same mass but different metallicity, the behavior of the  $r_{in}$ looks essentially the same.  The only differences are the evolutionary time-scales (the lower metallicity stars evolve more quickly) and the maximum value of $r_{in}$, which is a factor of 2 smaller for the most metal-poor model compared to the most metal rich.
In Figure \ref{fig:ainzoom}, we show the RGB evolution of four of the stellar mass models in Figure \ref{fig:ain} on a linear age scale to clarify trends for this phase of evolution
that occurs more rapidly.

\section{RESULTS}
\label{sec:results}
For the 99 PH stars of known luminosity class, we find that 89 stars  accrete at least one planet, 10 stars accrete no planets, and 36  become rapid rotators at some point  on the RGB  by gaining $\Delta$\vsini\ $\ge 8$~\kms.  The properties of these 36 star systems are listed in Table \ref{rrtable}, which gives
the name, mass and metallicity of the star, mass and semimajor axis of the accreted planet, the  maximum rotational velocity the star attains on the RGB, the evolution stage of the star when the planet is accreted (either MS, sub-giant (SG), or RGB), and the fraction of the RGB lifetime that the star is a rapid rotator ($l_{RR}$).   For the two stars that have two known companion planets,  HD 168443 b and HD 202206 b are accreted in our models while HD 168443 c and HD 202206 c are not.

\begin{deluxetable*}{lllrrrrr}
\tablewidth{0pc}
\tablecaption{Exoplanet host stars that are rapid rotator progenitors\label{rrtable}}
\tablehead{
\colhead{Star Name} & 
\colhead{$ M_{*}$} &
\colhead{[Fe/H]}&
\colhead{$ M_{p}\tablenotemark{a}$ } &
\colhead{$ a_{p}\tablenotemark{a}$}&
\colhead{Max. $\Delta v\sin i$} &
\colhead{$t_{in}$}&
\colhead{$l_{RR}$} \\
\colhead{} & 
\colhead{(\msun)} &
\colhead{(dex)} &
\colhead{($ M_{Jup}$)} &
\colhead{(AU)}&
\colhead{ (\kms)} &
\colhead{(Stage)} &
\colhead{(\%)}}
\startdata
\object{70 Vir}& 1.10&-0.03&  7.44& 0.48& 15.4  &RGB&2\\
\object{CoRoT-Exo-1}\tablenotemark{b} & 0.95&-0.30&  1.03& 0.03& 15.8&MS&14\\
\object{CoRoT-Exo-2}\tablenotemark{b} & 0.97&\ 0.00\tablenotemark{c}&  3.31& 0.03& 56.8&MS&70\\
\object{CoRoT-Exo-3}\tablenotemark{b} & 1.27&-0.05& 21.60& 0.05&286.3&MS&98\\
\object[HAT-P 2]{HAT-P-2}\tablenotemark{b} & 1.30&\ 0.11&  8.62& 0.07&115.1&SG&91\\
\object[HAT-P 5]{HAT-P-5}\tablenotemark{b} & 1.16&\ 0.24& 1.06& 0.04& 14.3&SG&15\\
\object[HAT-P 6]{HAT-P-6}\tablenotemark{b} & 1.29&-0.13&1.06& 0.05& 14.2&SG&17\\
\object[HAT-P 9]{HAT-P-9}\tablenotemark{b} & 1.28&\ 0.12& 0.78& 0.05& 10.2&SG&16\\
\object{HD 136118} & 1.24&-0.07& 11.90& 2.30&  9.3&RGB&0.06\\
\object{HD 141937} & 1.00&\ 0.16&  9.70& 1.52& 11.0&RGB&0.2\\
\object{HD 168443} & 1.06&\ 0.03&  8.02& 0.30& 27.3&RGB&11\\
\object{HD 178911 B}  & 1.07&\ 0.28&  6.29& 0.32& 14.5&RGB&3\\
\object{HD 179949}& 1.28&\ 0.22&  0.95& 0.05& 12.5&SG&16\\
\object{HD 202206}& 1.13&\ 0.37& 17.40& 0.83& 28.0&RGB&2\\
\object{HD 209458}\tablenotemark{b}& 1.01&\ 0.04& 0.69& 0.05&  9.9&RGB&12\\
\object{HD 33564} & 1.25&-0.12&9.10& 1.10&  9.3&RGB&0.1\\
\object{HD 68988} & 1.18&\ 0.24&  1.90& 0.07& 13.6&RGB&22\\
\object{HD 73256} & 1.24&\ 0.29&  1.87& 0.04& 26.1&SG&34\\ 
\object{HD 86081} & 1.21&\ 0.26& 1.50& 0.04& 20.7&SG&26\\  
\object{OGLE-TR-10}\tablenotemark{b} & 1.18&\ 0.12&  0.63& 0.04&  8.6&SG&15\\
\object{OGLE-TR-56}\tablenotemark{b} & 1.17&\ 0.00\tablenotemark{c}&1.29& 0.02& 12.0&MS&16\\
\object{OGLE-TR-132}\tablenotemark{b} & 1.26&\ 0.37&  1.14& 0.03& 11.5&MS&16\\
\object{OGLE-TR-182}\tablenotemark{b} & 1.14&\ 0.37&  1.01& 0.05& 17.5&SG&16\\
\object{SWEEPS-04}\tablenotemark{b} & 1.24&\ 0.00\tablenotemark{c}& 3.80& 0.06& 63.1&SG&75\\
\object{Tau Boo}& 1.30&\ 0.28&  3.90& 0.05& 51.5&SG&71\\
\object{TrES-2}\tablenotemark{b} & 0.98&-0.15&  1.20& 0.04& 23.9&MS&36\\
\object{TrES-3}\tablenotemark{b} & 0.92&-0.19&  1.92& 0.02& 30.7&SG&41\\
\object{WASP-1}\tablenotemark{b} & 1.24&\ 0.00\tablenotemark{c}&  0.89& 0.04& 12.1&SG&16\\
\object{WASP-3}\tablenotemark{b} & 1.24 &\ 0.00\tablenotemark{c} & 1.76 & 0.03 & 20.7&MS&26 \\
\object{WASP-5}\tablenotemark{b} & 0.97 &\ 0.00\tablenotemark{c} & 1.58 & 0.03 & 27.1&MS&36\\
\object{WASP-14}\tablenotemark{b} & 1.32&\ 0.00\tablenotemark{c}& 7.73& 0.04& 93.0&MS&88\\
\object{XO-1}\tablenotemark{b}& 1.00&\ 0.00\tablenotemark{c}& 0.90& 0.05& 20.3&RGB&27\\
\object{XO-2}\tablenotemark{b}& 0.98&\ 0.45&  0.57& 0.04& 11.0&SG&17\\
\object{XO-3}\tablenotemark{b}& 1.21&-0.18& 11.79& 0.05&167.7&MS&93\\
\object{XO-4}\tablenotemark{b}& 1.32&-0.04&  1.72& 0.06& 25.5&SG&36\\
\object{XO-5}\tablenotemark{b}& 1.00&\ 0.18&  1.15& 0.05& 25.0& SG& 36 \\
\enddata
\tablenotetext{a}{In all cases, the  accreted planet was the first discovered for that star, e.g., HD 168442 b  was accreted but HD 168443 c was not.} 

\tablenotetext{b}{Transiting system} 

\tablenotetext{c}{No measured stellar [Fe/H]} 

\end{deluxetable*}

\begin{figure}
\epsscale{1.0}
\plotone{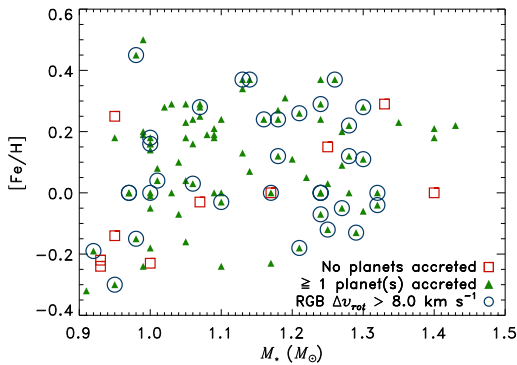}
\caption{Distribution of planet hosting stars that will not accrete any planets (squares), that will  accrete at least one planet (triangles), and that will be spun-up enough by the accretion of one or more planets to become rapid rotators for part of their red giant lifetime (circles). }
\label{fig:smass}
\end{figure}

\begin{figure}
\epsscale{1.0}
\plotone{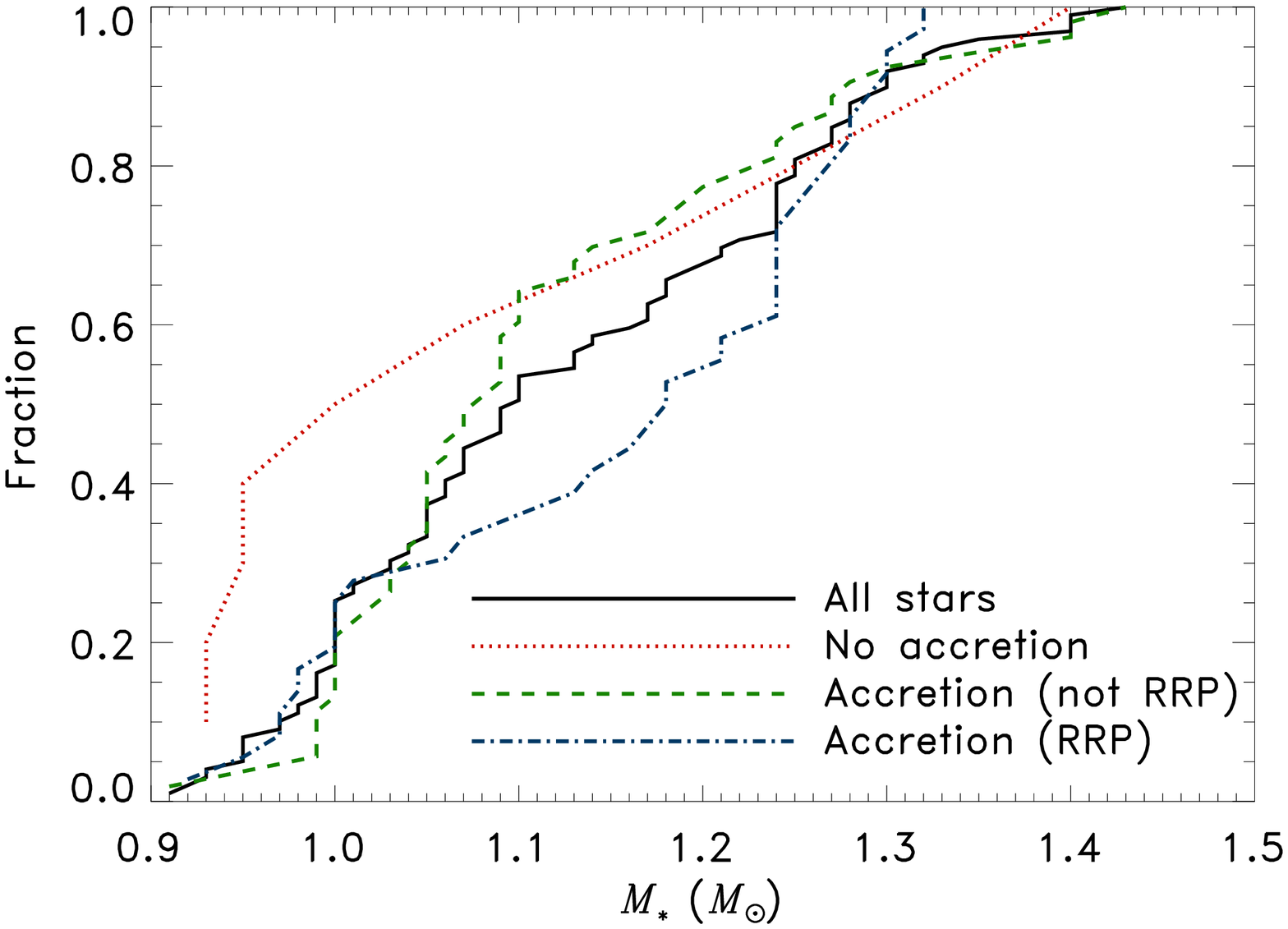}
\caption{Discrete CDF of the stellar masses of: all stars in the sample (solid), PH stars that will not accrete any planets (dotted), PH stars that will accrete planets but not become rapid rotators (dashed), and PH stars that will be rapid rotators (dot-dashed).}
\label{fig:hsmass}
\end{figure}

\begin{figure}
\epsscale{1.0}
\plotone{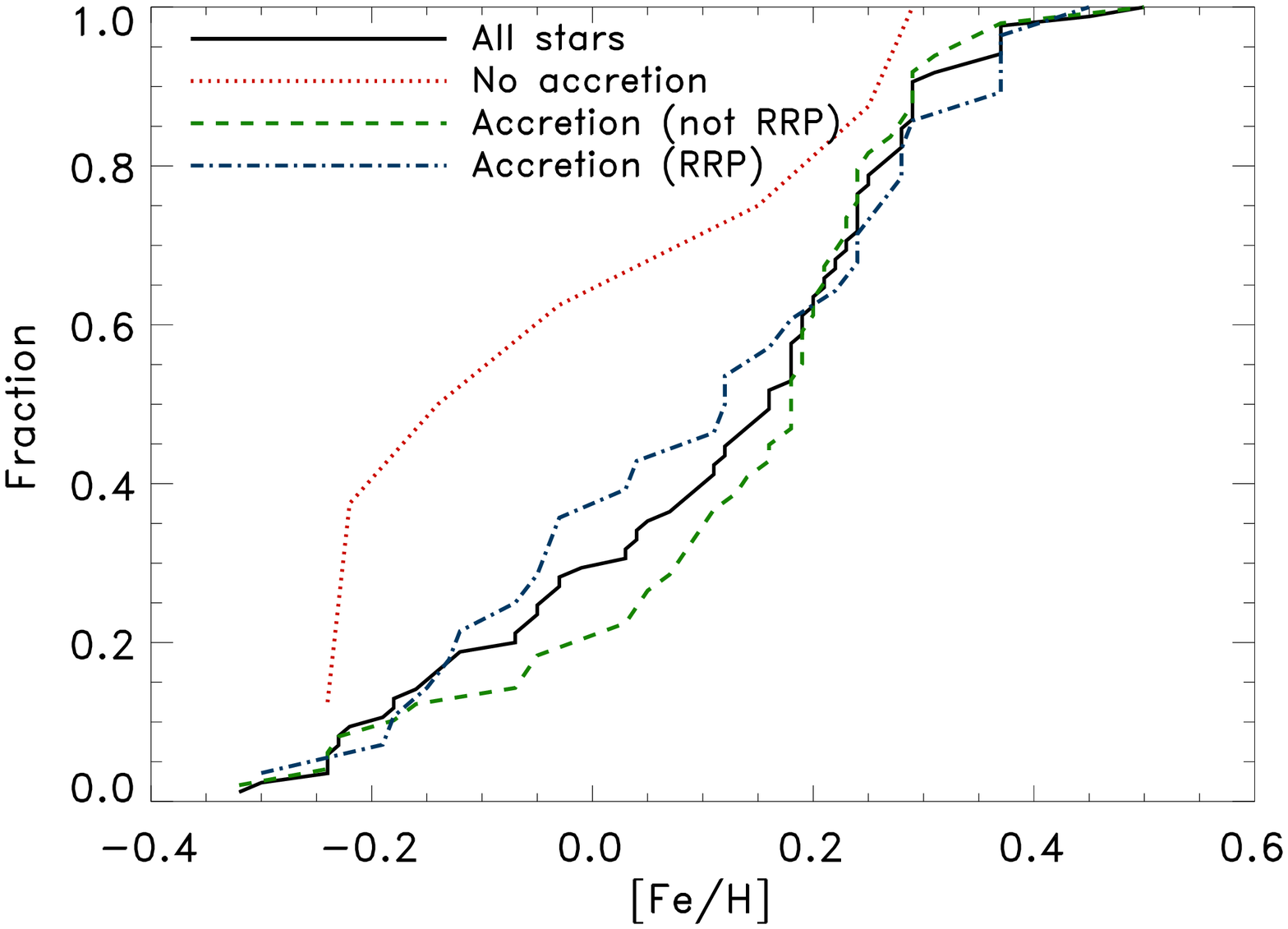}
\caption{ Discrete CDF of the metallicities of: all stars in the sample (solid), PH stars that will not accrete any planets (dotted), PH stars that will accrete planets but not become rapid rotators (dashed), and PH stars that will be rapid rotators (dot-dashed). }
\label{fig:hfeh}
\end{figure}

In Figure \ref{fig:smass}, we compare $M_*$ and [Fe/H] of the planet-hosting stars that accrete planets, that do not accrete planets, and that are red giant rapid rotator progenitors  (RRPs) to see whether any trends exist.  In general, there are no strong correlations between $M_*$ or [Fe/H] and whether a PH star absorbs a planet.   To better explore the differences in these samples, we plot cumulative distribution functions  (CDFs) in $M_*$ and [Fe/H]  in Figures \ref{fig:hsmass} and \ref{fig:hfeh}, respectively.  Plotted separately are CDFs of the entire PH stellar sample, PH stars that do not accrete planets, PH stars that accrete one or more planets but are not RRPs, and the RRPs.
The distributions of stars that accrete planets diverge from  the distribution of all the stars in general in the mass range of 1.05-1.25~\msun.   In this mass range, a larger fraction of the stars that  accrete planets do not become rapid rotators. 
The two-sided Kolmogorov--Smirnov (KS) statistic reveals that the probability that  the RRPs and the stars that ingest planets but are not RRPs are drawn from the same parent population 
is only 4\%.  The stars that do not accrete planets, which is the smallest subsample  comprises  a larger fraction of  low-mass stars than the parent sample. 
Comparing all stars that ingest planets to all those that do not, the KS probability of being from the same parent sample is  16\%.

\begin{figure}
\epsscale{1.0}
\plotone{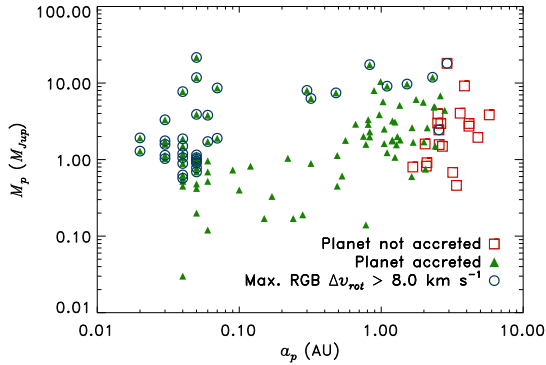}
\caption{Distribution of planet mass and semimajor axis of exoplanets that will survive (squares), exoplanets that will be accreted  (triangles), and exoplanets orbiting a star that is a RRP (circles). Because some planets are in multiple systems, a planet that will survive might still orbit an RRP. }
\label{fig:pmau}
\end{figure}

\begin{figure}
\epsscale{1.0}
\plotone{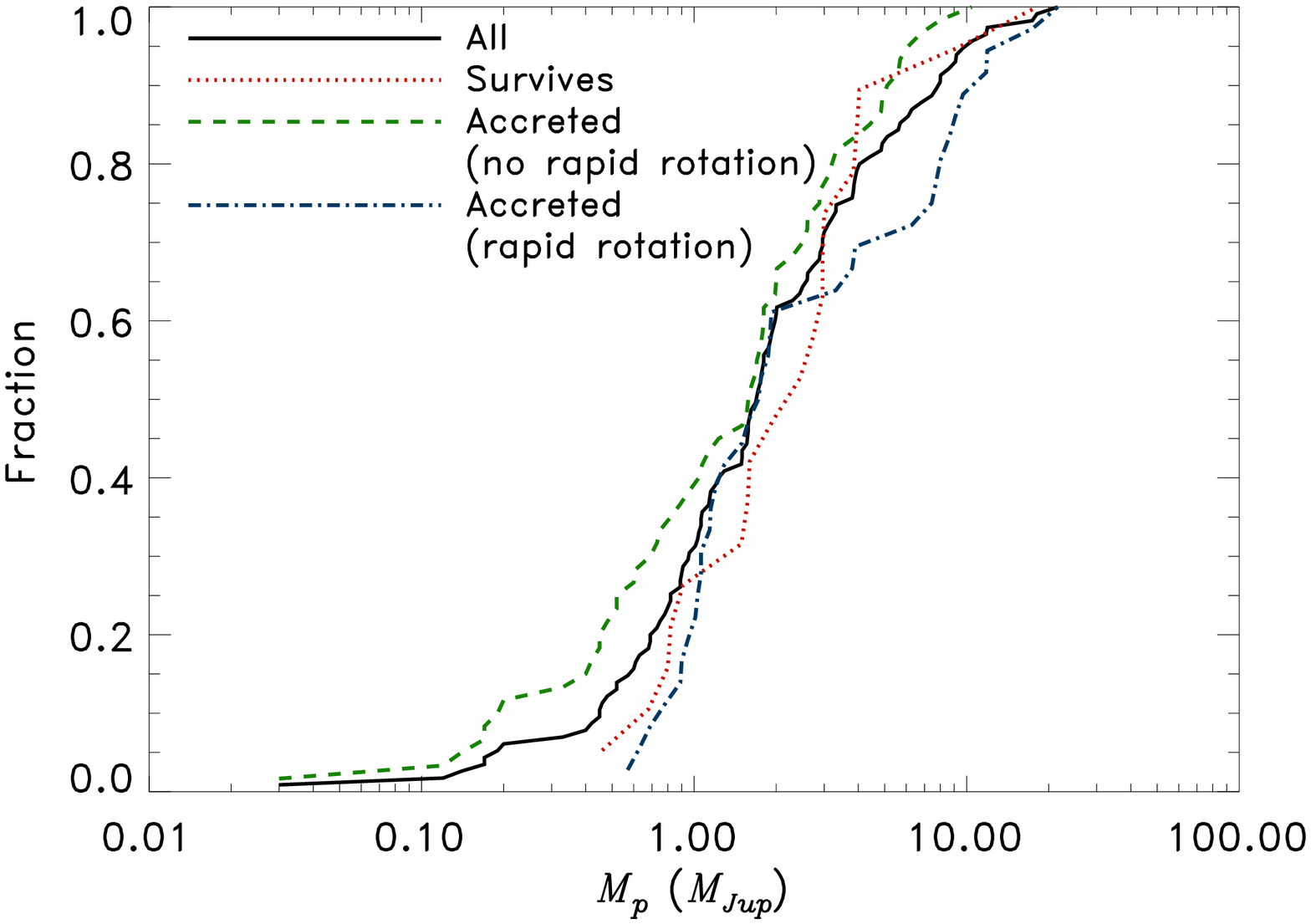}
\caption{Discrete CDF of the planetary masses of: all the exoplanets  (solid), exoplanets that will survive (dotted), exoplanets that will be  accreted  but not create rapid rotation (dashed), and exoplanets that will be accreted orbiting RRPs (dot-dashed). }
\label{fig:hpm}
\end{figure}

\begin{figure}
\epsscale{1.0}
\plotone{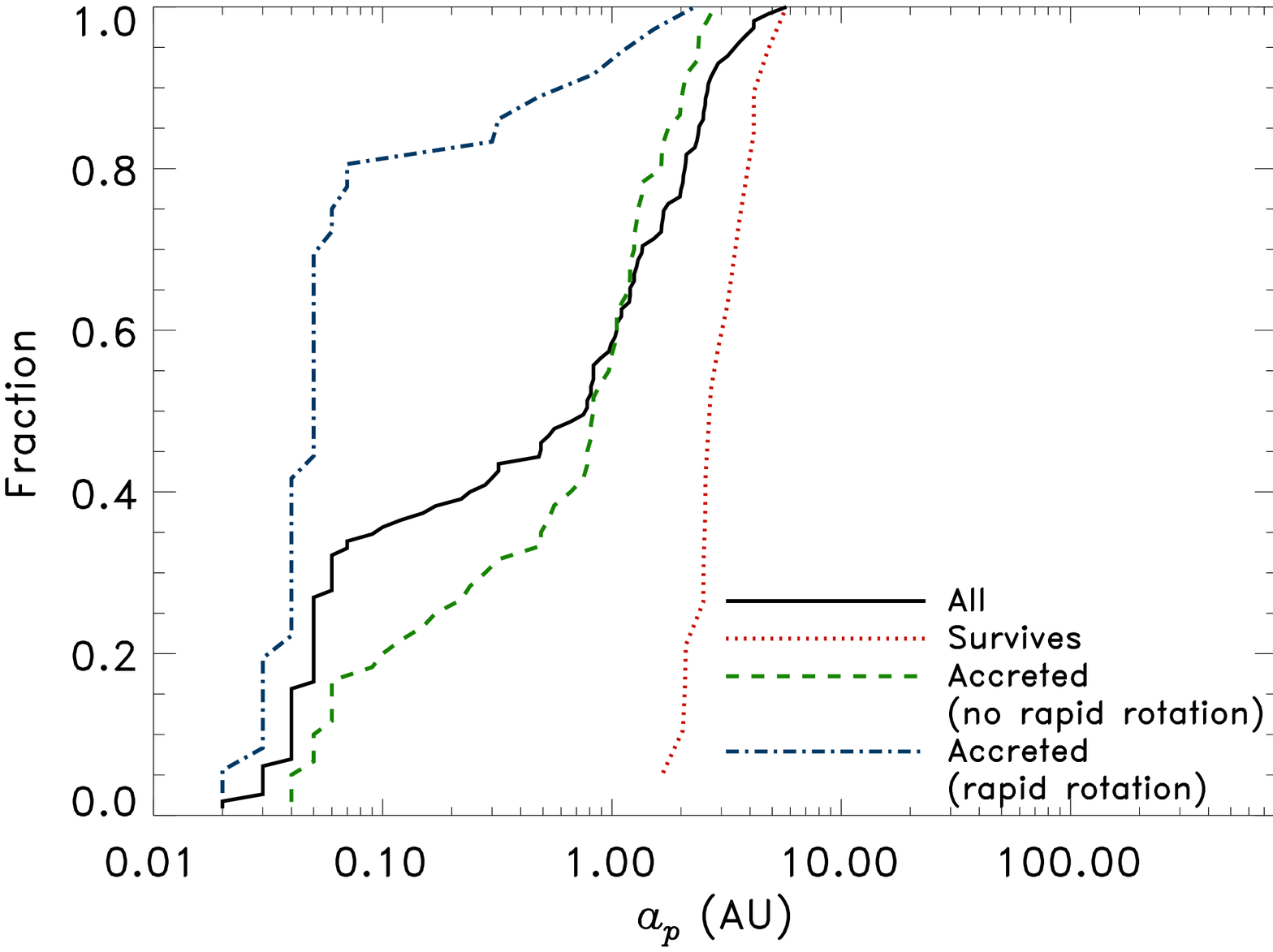}
\caption{Discrete CDF of the semimajor axes of:  all the exoplanets  (solid), exoplanets that will survive (dotted), exoplanets that will be  accreted  but not create rapid rotation (dashed), and exoplanets that will be accreted orbiting RRPs (dot-dashed). }
\label{fig:hau}
\end{figure}

The [Fe/H] CDFs in Figure \ref{fig:hfeh} show that  the low [Fe/H] stars tend not to accrete planets as readily. For this plot, we removed all stars  for which we had assumed solar [Fe/H]. Interestingly, this plot also shows that of the stars that do ingest planets, those that are RRPs are somewhat more metal poor than those that are  {\it not} RRPs. Repeating the KS-test 
for the [Fe/H] distribution, we find that there is a 54\% chance that the RRPs and the stars that ingest planets but are not RRPs are from the same parent population.  Stars that do not ingest planets and stars
that do ingest planets have  a 19\% chance of being drawn from the same population.   The stars that do not ingest planets form the most metal poor subsample.
A tentative conclusion is that both the $M_*$ and [Fe/H] are correlated with the probability of a PH star accreting a planet and becoming a rapid rotator, albeit weakly.  The metal poor PH stars are less likely to accrete their planets, but of those that do accrete planets, they are more likely to be RRPs.   Stars slightly more massive than the Sun that accrete planets are correlated with a lower probability of becoming
a rapid rotator, although it should be noted that this may be a reflection of the [Fe/H] effect; the stars in this mass range are on average more metal rich than the sample outside that range ---0.13 dex compared to 0.08 dex (excluding stars with [Fe/H] set to solar in both cases).

The role of the planetary parameters in  planet accretion and the creation of rapid rotators is illustrated in Figures \ref{fig:pmau}--\ref{fig:hau}.  Figure \ref{fig:pmau} shows the distribution of $M_p$ and $a_p$ for the planets that survive the entire RGB evolution, planets that are accreted, and planets with host stars that are RRPs.  Note that there are two planets that survive the RGB phase  orbiting an RRP that is spun up by the accretion of a closer-orbiting planet. Most planets with $a_p < 2$~AU are accreted.  CDFs of all the planets, planets that survive, planets that are accreted but do not cause rapid rotation, and planets that do create rapid rotation are plotted for  $M_p$ (Figure \ref{fig:hpm}) and $a_p$ (Figure \ref{fig:hau}).  Figure \ref{fig:hpm} reveals  that no planet with $M_p \lesssim 0.5~M_{Jup}$ has enough angular momentum to create a rapid rotator.   The CDFs in $a_p$ of these different samples in Figure \ref{fig:hau} show large disparity.  As expected,  planets that are ingested generally have smaller semimajor axes than those that are not ingested.  
However, this figure also shows that the ingested planets that orbit RRPs have a distribution more heavily weighted to {\it smaller} semimajor axes compared to the sample of ingested planets that do not orbit RRPs, despite the fact that more distant planets have more angular momentum for a given planet mass.  The KS test gives a 3\% probability of these two samples being drawn from the same parent population.    
This result is explained by the fact that the tidal decay timescale depends  on $(a_p/R_*)^8$.  The planet spirals into the star
essentially  when $R_*$ reaches roughly some critical fraction of $a_p$.  Consequently, while the available orbital angular momentum is increasing with $\sqrt a_p$, the total $\Delta v_{rot}$ added to the star is decreasing with $\sqrt a_p$ (i.e.,  the behavior of Equation (\ref{eqvrot}) for $R_*\propto a_p$).  The  spin-up effect from the larger angular momentum available in planets with larger orbital radii is more than offset by the increased stellar moment of inertia when these outer planets are finally accreted, and we see this effect in Figure \ref{fig:hau}.

\begin{figure}
\epsscale{1.0}
\plotone{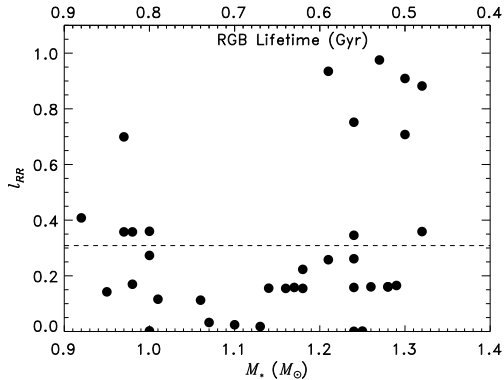}
\caption{Fraction of the RGB lifetime that the planet-hosting rapid rotator progenitor  will spend as a rapid rotator, $l_{RR}$, as a function of the stellar mass.  Approximate RGB lifetimes are given on the top axis.
The dashed line indicates the mean value of $l_{RR}$.}
\label{rrprog}
\end{figure}

In order to compare these results to the Galactic red giant population in general,  we must have a sense of  not only how many PH stars become rapid rotators but also how long they remain rapid rotators.  
Figure \ref{rrprog} shows the rapid rotation lifetime, $l_{RR}$, as a fraction of the total RGB lifetime,  for the 36 RRPs as a function of the model $M_*$; the approximate  total RGB lifetime is given on the top axis.   The average value of $l_{RR}$ is 31\%~$\pm$~4\%, and this is indicated by the horizontal line.  Although there  is a  slight positive correlation between the stellar mass and $l_{RR}$, there is also significant scatter in $l_{RR}$, particularly at the highest  and lowest masses ($M_*>1.2$\msun\ and $M_*<1.0$\msun).  Thus, we assume that the average fractional lifetime is applicable to a generic population of red giants  over a range of masses.

The expected fraction of rapid rotators, $f_{exp}$, predicted from our planet ingestion models can be written as $f_{exp}=f_{PH} f_{RRP} l_{RR}$, where $f_{PH}$ is the fraction of stars that host planets,
$f_{RRP}$ is the fraction of planet-hosting stars that will become rapid rotators from planet ingestion, and $l_{RR}$ is described above.
While the total fraction of stars that harbor planets of any mass or at any orbital distance is unknown, it is possible to calculate the fraction of stars with planets within the observational constraints imposed by exoplanet survey methods.  In particular,   \cite{grether06} find that 5\%$\pm$2\% of {\it all} Sun-like stars have at least one planetary companion orbiting within 3 AU.
Using $f_{PH}=0.05$, we calculate $f_{RRP}$ only considering those stars in our sample with at least one planet within 3 AU.  In our sample, the number of PH stars meeting this criterion, which we 
call $N_{PH}$, is  96, and of these stars, 36 accrete a planet  with sufficient orbital angular momentum early enough in their evolution to cause rapid rotation.
Therefore,  the fraction of RRPs, $f_{RRP}$,  in this sample is  37.5\%.  Assuming the stars spend an average  of 31\% of the RGB lifetime as rapid rotators, then the expected fraction of rapid rotators, $f_{exp}$, among single red giants from the accretion of planetary companions is 
0.58\%$\pm$0.26\%.  The actual frequency of rapid rotation in various surveys of red giants ranges from 2-5\%. As such,  planet accretion can account for only 10-30\% of these stars {\it if} the present known PH sample is a reasonable proxy for field RGB progenitors.  These results ($N_{PH}$, $f_{RRP}$,   $l_{RR}$, and $f_{exp}$) are tabulated in Table \ref{tab:results} under ``Case A.'' 
\begin{deluxetable}{lrrrr}
\tablewidth{0pc}
\tablecaption{Summary of results for subsets of  the PH stars. \label{tab:results}}
\tablehead{
  \colhead{} &
  \colhead{$N_{PH}$ ($a_p < 3$ AU) } &
   \colhead{$f_{RRP}$} &
   \colhead{$\overline{l_{RR}}$} &
   \colhead{$f_{exp}$ } }
\startdata
Case A & 96 &37.5\% & 31\%& 0.58\% \\
Case B &  130 &33.8\% & 31\%& 0.53\% \\
Case C &  96 & 33.3\% & 33\% & 0.55\%
\enddata
\end{deluxetable}

If we include the stars with unknown luminosity class and assume that they are likely on the main sequence, we add an additional 34 stars with at least one planet orbiting within 3 AU,  of which 31 ingest planets, and 8 of these are RRPs.  Following the same calculations described in the previous paragraph, we find  0.53\%$\pm$0.23\%  of low-mass red giants should be rapidly rotating due to planet ingestion; these results are summarized under ``Case B'' in Table \ref{tab:results}.

\section{DISCUSSION}
\label{sec:discuss}

\subsection{Evaporating Giant Planets in the Stellar Envelope}
\label{sec:evap}
In our models, the issue of whether a planet is fully evaporated when it is accreted can affect the amount of angular momentum deposited by about a factor of 2.    As noted earlier, 50\% of the planet's
angular momentum is added to the star simply from tidal decay of its orbit  from the initial separation to the stellar surface.  We assume the remaining angular momentum is contributed to the star through
complete subsumption, and this assumption warrants further discussion.
 A complete physical treatment of the planet evaporation would result in a significant  increase not only in the complexity of the physics in our models (particularly since we are modeling almost 100 individual systems) but also in the number of assumptions regarding unknown characteristics of these exoplanets, such as their physical radii, compositions, and density profiles.   Instead, we look to  previous models of the evaporation of planets in main sequence stars  \citep{sandquist98,sandquist02} and giant stars \citep{livio84,soker98} for some insight.

For MS stars, \cite{sandquist98} modeled the accretion of Jupiter- and Saturn-like planets into stars of 1~\msun\ and 1.22~\msun\ and found that even though the two stellar models differed by only 0.22~\msun, the details of the stellar structure were different enough to change the amount of evaporation from total (1~\msun\ case) to only one-third of the  original planet mass  (1.22~\msun\ case).
Additionally,  they found a sensitive dependence of the amount of evaporation on the structure of the planetary interior; planets with steeper density gradients had higher survival rates than those with shallower density gradients.  For giant stars, \cite{livio84} found that planets
that spiraled into their stars could suffer one of three fates --- complete evaporation, partial evaporation, or mass accretion.  Their models covered a range of planet masses, initial separations, and physical assumptions for stellar models at  0.88 and 1.2~\msun.   They found that a critical planet mass, $m_{crit}$, exists above which a planet will accrete mass and below which a planet will evaporate.
Depending on the model and physical assumptions, this critical mass was found to lie between  $\sim$7.5 and 10~$M_{Jup}$. For the models that gave the smaller value, planets up to only
$\sim$5~$M_{Jup}$ are evaporated completely, while those with masses between 5 and 7.5 $M_{Jup}$ are partially evaporated.   However, a later paper by \cite{siess99} gives  $m_{crit}\sim20~M_{Jup}$.
For stars near the RGB tip (which have the most rarified atmospheres), \cite{soker98} found a relationship between  the fate of planets and the planet mass.  Planets of order a Jupiter mass or less evaporate at distances from the stellar center that depends on the planet mass. A Jupiter mass planet evaporates  at approximately 10~\rsun, and less massive planets evaporate at larger radii.  Planets more massive than Jupiter may avoid evaporation until they are deep enough in the star to overflow their  Roche lobes; therefore, the most massive destroyed planets may contribute angular momentum to the core rather than the envelope. 

In light of all the uncertainties in these models, we conclude that our assumption of complete evaporation is not unreasonable, particularly for the lower mass planets.  
If we take a more conservative stance and assume that planets more massive than $5~M_{Jup}$ do not evaporate, then these planets only add angular momentum to their stars that is equal to  the difference between the orbital angular momentum when the planet is at $a_p$ and when the planet is at $R_*$.  Applying this assumption to the PH stars we modeled in Case A, we find that four of the stars will no longer become rapid rotators; therefore, $f_{RRP}$ drops to  33\%.  Additionally,  the average lifetime increases slightly  to 33\%, and $f_{exp}$ becomes 0.55\% $\pm$ 0.25\%.   These results are listed under ``Case C'' in Table \ref{tab:results}. 

\subsection{Effects of Mass Loss on the RGB}
\label{sec:mdot}
\begin{figure*}
\epsscale{0.67}
\plotone{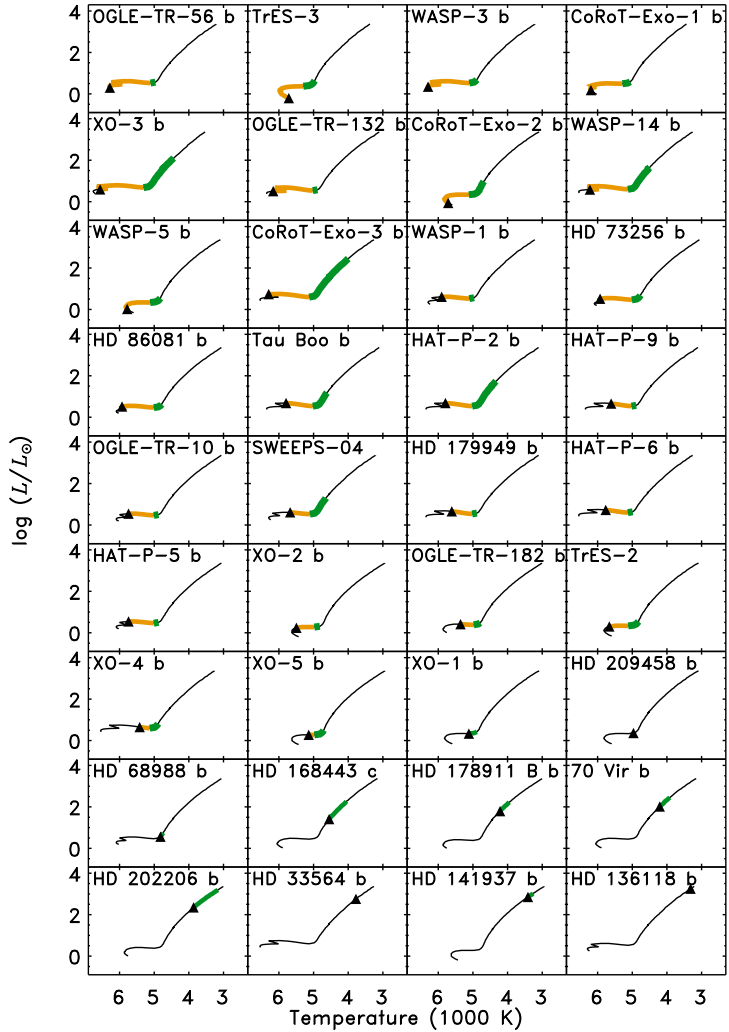}
\caption{Stellar evolution tracks on the HR diagram for the RRPs.  The triangles mark when the planets are ingested. Thick colored regions of the tracks mark rapid rotation that
occurs before the RGB base in orange and from the RGB base onward in green. 
Panels are ordered by time of planet ingestion from early to late in the stellar  evolution.}
\label{fig:eathr1}
\end{figure*}

Ignoring mass loss will not affect the number of possible giant rapid rotators if the stars accrete planets on the RGB itself; it only affects the time that star spends as a rapid rotator.  However, a surprising result of our simulation is the number of planets for which ingestion occurs before the RGB. Figure \ref{fig:eathr1} shows on the Hertzsprung--Russell (HR) diagram the evolution of the PH stars that will become rapid rotators.  The triangles mark  the evolution stage when the planets are ingested.  Thick regions on the HR diagram indicate when $\Delta$\vsini$\ge$ 8~\kms.  Of these RRPs, we find that 26 (72\%) accrete planets before reaching the base of the RGB, and  10  (28\%) do so while still on the main sequence.    Of all the PH stars that ingest planets, 28  (31\%) do so before arriving at the base of the RGB  and 10  (11\%)  ingest planets before leaving the main sequence.
All of the stars that ingest planets on the MS are transiting planets, a result consistent with  the recent evidence that many transiting planets are in the midst of tidal migration and are not in stable orbits \citep{Jackson08}.  Such tidally evolved planets are more likely to be found with transiting surveys because the range of inclination angles  resulting in a transit increases for closer orbiting planets.

The accretion of planets before the RGB is a significant result  because these pre-RGB stars have yet to undergo rotational braking \citep{Gray89}, which is a magnetic dynamo mechanism thought to be responsible for the nearly ubiquitous low rotational velocities of red giants.  In Figure \ref{fig:eathr1}, this transition occurs at a temperature of $\sim$5550~K. 
The proposed braking mechanism is triggered by the deepening convective envelope;  the associated magnetic field forces co-rotation of escaping mass, allowing that mass to carry away large amounts of angular momentum. Thus, if planetary angular momentum is added to the star prior to this braking mechanism, that angular momentum will be lost before the star reaches the RGB. In addition, this braking mechanism  should continue working throughout the RGB evolution, decreasing the amount of time that the rapid rotators spend in this high rotation state while on the RGB.    Considering this diminishing effect of magnetic braking, planet ingestion probably accounts for less of the rapid rotator population than found in \S \ref{sec:results}.

Even in the absence of a magnetic dynamo, significant angular momentum can be carried away by stellar winds as demonstrated by
\cite{soker00}, who studied the angular momentum evolution of  spun-up RGB stars onto the HB. 
The rate that  stellar angular momentum is  lost ($\dot{J}_*$) is proportional to the stellar angular velocity ($\omega_*$), the mass loss rate ($\dot{M}$), and $R_*^2$.
Ignoring interactions with companions for the moment, we expect the angular momentum loss to increase as the star ascends the RGB because of mass loss.  
Although the \cite{giard00} evolution models we used ignore mass loss,  models that do include mass loss \citep[e.g.,][]{charbonnel99} show that mass loss accelerates as a star ascends the RGB and is thus most substantial near the tip. 
Additional angular momentum is lost when one considers the interactions with companions. The tidal interactions that cause planets to spiral in will increase $\omega_*$, and
\cite{siess99} found that planet absorption can actually {\it increase}  $\dot{M}$ on the RGB.  Both of these will increase $\dot{J}$. Consequently, ignoring mass loss will lead to increasingly inaccurate results the further along the stellar evolution the assumption  is applied, which is why we chose to end our simulations at the RGB tip. 

To further illustrate the importance  of considering mass loss in an analysis of planet-induced rotation in post-MS stars, we analyze the recent work by
\cite{massarotti08b}. Contrary to our results, \cite{massarotti08b} finds that HB stars have the highest probability among evolved stars of exhibiting detectably larger rotational velocities due to planet ingestion.
However, this result is only possible if  the star can hold onto a substantial fraction of its envelope mass and any added angular momentum for the entire RGB evolution through the helium flash. 
Under this assumption, a star at the RGB tip can ingest planets with relatively large $a_p$ (and  thus  relatively larger angular momentum) and {\it not} become a rapid rotator,  but when that same star evolves to the HB, the stellar radius and moment of inertia decrease enough for that same angular momentum to cause rapid rotation.  However, given that a significant fraction of the accreted angular momentum may be lost, the probabilities derived for the HB stars may in fact be much lower than calculated in \cite{massarotti08b}. Moreover, the role of planet accretion by  RGB stars on the nature of HB stars is probably far more complicated than a simple calculation of rotational velocity from the net change in the stellar angular momentum.  
The degree of mass loss, particularly driven by planet accretion, \citep[][and references therein]{pete83,soker00}, and the redistribution of angular momentum between the stellar core and envelope during helium flash  \citep{soker07}
has been proposed  to be responsible for a rich variety of HB star properties, e.g., temperature, rotation, and luminosity,  that cannot be explained by other means.

\subsection{PH Stars as a Proxy for  Red Giant Populations}
\label{sec:feh}
In this section, we discuss the suitability of using main sequence PH stars as a representative  sample of  the precursors to present-day  red giant stars.  While these PH stars are the obvious choice for investigating the role of planet accretion in spinning up giant stars, they are not an ideal choice. Obviously red giants are further along in their stellar evolution than the PH stars, which implies that the giants are either older on average than the PH stars or they are of similar age but  have evolved faster than the PH stars.   Faster evolution occurs in stars that are relatively more massive or more metal poor. 
Among the global properties of the PH stars and the red giants that may differ in a significant way, we focus on the metallicity, mass,  and spatial distributions. 

Stellar metallicity complicates the interpretation of our results because of the well-defined stellar metallicity-planet frequency  (MPF) correlation \citep{fischer05} for main sequence stars, which is
the power-law relationship between the metallicity of a MS star and the likelihood that the star has a planetary companion.  The \cite{sozzetti09} planet hunting survey of low metallicity stars extended the metallicity range of the correlation down to -2.0 dex.  Although they confirmed the power law behavior in the more metal-rich stars, they also found some indication that the power law may flatten to a constant probability at low metallicities.
Whether this MPF correlation is  ``nature'' or ``nurture'' is disputed.  Either metal-rich stars are predisposed to having planets (nature) or the presence of planets can enrich the apparent stellar atmospheric metallicity (nurture).

In support of the nature scenario,  \cite{fischer05} found that there was no trend of stellar metallicity with stellar temperature in main sequence PH stars.  This result is evidence for  primordial metal-richness of PH stars because hotter MS stars have shallower convective zones;  if the atmospheres of PH stars were enriched with metals through the planet formation process, then the hotter stars with shallower envelopes would have higher apparent metallicities than cooler stars with deeper envelopes. Supporting the nurture explanation is the \cite{pasquini07} discovery that PH {\it giant} stars do not show a MPF correlation. The most natural explanation for the absence of the correlation is that the significantly deeper convective envelopes of giant stars have erased the planet formation pollution seen in the main sequence PH stars.  One would expect to see the MPF correlation in giant stars if intrinsically metal rich stars are predisposed to forming planets.

Either explanation for the MPF correlation has some relevance to our analysis and the interpretation of our results.
 A comparison of the metallicities of the PH stars in Figure \ref{fig:smass}  with the red giants  in Figure \ref{fig:smassfeh} shows that the PH stars are on average more metal rich than the giants.  Under the nature scenario, we consequently overestimate the original planet frequency in the present metal-poor giant populations and overestimate the contribution of planet ingestion to the rapid rotation seen in these stars.  
Conversely,  under the nurture scenario,  the primordial metallicities of the PH stars are lower than those presented in Figure \ref{fig:smass} and more closely match those of the giants.  However, if only the thin atmospheres of PH stars are polluted with metal-rich material,
then we modeled the evolution with stellar models that were too metal-rich.  As an experiment  to see how this affects the outcome of our simulation, we reran our simulations, matching each PH star with stellar models that were systematically more metal-poor than the measured [Fe/H].  We tried a range of systematic offsets between -0.1 and -2.0 dex in increments of 0.1 dex.   We find that the total number of RRPs in our simulation varied over the range of 36 to 42, while the average lifetime spent as a rapid rotator steadily dropped from 31.0\% down to 18.4\%. 
Therefore, our possible overestimation of the stellar metallicity because of pollution may have also resulted in an overestimate of the number of expected rapid rotators from planet ingestion.

The differences in the $M_*$ distributions of the PH stars and red giants are also important. As illustrated in Figure \ref{fig:smassfeh}, the actual mass range covered by a sample of field red giant stars extends to higher masses than what is probed by the most  prolific exoplanet surveys. This is primarily a bias in the exoplanet surveys.  More massive main sequence stars have broader spectral features and more atmospheric ``jitter," i.e., larger amplitudes in the random variations of the measured radial velocities originating in the stellar atmosphere.  Both of these effects severely limit the attainable radial velocity precision that is needed to detect the periodic stellar wobbles that imply the presence of a planet. \cite{Johnson06} began a survey to circumvent these difficulties by probing the planet frequency around more evolved massive stars.  In particular, his survey focuses on the sub giant branch where masses are most easily discriminated on a color magnitude diagram.  Preliminary results from this survey point toward a paucity of close-orbiting planets \citep{Johnson07, Johnson08}; none of the 15 planets discovered in this survey orbit closer than 0.8 AU.  This observed paucity cannot be an observational bias because radial velocity surveys are most sensitive to the {\it smallest} orbital  semimajor axes.
Nor is this inner limit consistent with the maximum radius of planets that have tidally spiraled into the star; our predictions from Figure \ref{fig:ain} suggest an inner limit that is at least an order of magnitude smaller. \cite{Johnson07} also considered post-MS engulfment of planets to be unlikely.  Presently, the most plausible explanation is that the lack of close-orbiting planets around more massive stars is a natural consequence of planet formation and migration and gives an indication the role of $M_*$ in these two processes \citep{Johnson08}.
 In our simulation, of the 50 stars harboring planets no closer than  0.8 AU, only 4 (8\%)  are predicted to be RRPs.  Consequently, these more massive stars might have a value of  $f_{RRP}$ that is four times lower than that of the stars we probed.   On the other hand, $M_*$ also plays a role in the planet frequency, and stars more massive than  1.3\msun\ are five times more likely to harbor planets than low-mass M dwarfs \citep{Johnson07b}, which more than compensates for the decreased efficiency of spin-up.  

Finally, differences in the spatial distributions of the PH stars and red giants may also affect the appropriateness of comparing these two stellar samples because of the possibility that 
the locally-observed planet frequency is not ubiquitous throughout the Galaxy.  Most known exoplanets were found by the radial velocity technique, and these surveys target nearby stars (within 100 pc), whereas the rapidly rotating field giants that we seek to explain are generally a few kpc away. One project that has made strides in addressing the deficiency of known extrasolar systems around distant stars is the Sagittarius Window  ExoPlanet  Survey \citep[SWEEPS,][]{sahu07}, which surveyed the Galactic bulge for one week using HST.
The SWEEPS survey found that the frequency of planets around stars in the bulge was consistent with those of nearby stars \citep{sahu07}, although the authors admit that their small sample size of only 16 planet candidates allows errors  of factors 2--3 in the planet frequency.  
Additionally, the bulge is relatively more metal rich than the solar neighborhood and perhaps significantly more so than many of the distant red giants comprising the present day rapid rotators.

 \subsection{Accretion Complements  Other Spin-up Mechanisms}
 \label{sec:other}
 Angular momentum dredge-up from the stellar core is an oft-cited possible alternative mechanism for spinning up red giant stars.   If giant stars are not rigid rotators and if their  cores have high specific angular momentum, then the deep mixing that occurs at first dredge-up can spin up giants' convective envelopes.  Rapid rotation in giant stars is only consistent with this mechanism if they have evolved beyond first dredge-up, whereas we have found that planet ingestion is most effective at creating rapid rotation in the earliest states of the RGB (recall Figures \ref{fig:vrotpot} and \ref{fig:eathr1}). Unfortunately, statistics of the observed occurrence of rapid rotation as a function of RGB evolution time are sparse.  Furthermore, any such statistics are complicated by the possibility of confusing RGB stars with stars in the red clump (the compressed HB of metal-rich stars) or AGB; all of these evolutionary stages overlap in an HR diagram particularly when the sample of giant stars have a wide range of ages, masses, and metallicities.  Nonetheless, it is reasonable to assume that rapid rotation on the lower RGB is likely due to planet accretion whereas rapid rotation on the upper RGB is more likely to be from angular momentum dredge-up because these mechanisms are most efficient at complementary locations on the RGB. 	
 \cite{massarotti08a} found three field giants in their survey that have moderately  high rotation, that are first ascent red giants that have not reached first dredge-up, and that they concluded could only be explained by planet accretion.  Of the single giants in their sample, these three giants that likely accreted a planet account for 0.4\% of their survey, and account for a slightly larger percentage of first ascent RGB stars.   This percentage is consistent with our prediction  that one would expect to find 0.58\% of a sample of first ascent red giants to have been spun up by planet accretion.
 
Alternatively, the disparity between the predicted number of rapid rotators from planet accretion and the larger fraction of rapid rotators seen in present-day giants may simply be resolved by the existence of
 low-mass stellar companions.  Although close-orbiting brown dwarf companions are rare, close-orbiting stellar companions occur in roughly 11\% of stars \citep{grether06}.  Consequently, 
some of the supposedly single red giant rapid rotators may in fact have undetected stellar companions with which  they have tidally interacted. Those stellar companions on the smallest mass scales are the most likely to avoid detection given their relatively low luminosity and gravitational effects on their hosts compared to more massive companions.

\section{SUMMARY}
This study was motivated by the question of whether the phenomenon of rapidly rotating red giant stars can be explained solely by spin-up from the accretion of planets. We chose the population of main sequence planet-hosting stars as a proxy sample of rapid rotator progenitors. Depending on the role of metallicity in the true frequency of planets among main sequences stars, our choice of a relatively metal-rich proxy biases us to deriving an {\it upper limit} to the number of rapid rotators generated by planet absorption among the local Galactic distribution of red giants.  On the other hand, other factors in our model, such as our possible overestimation of the convective envelope mass beyond first dredge-up described in \S \ref{sec:evmod} and the choice of a relatively large value for deviant rotation rates, i.e., \vsini~$\ge$~10~\kms\ as deviant in stars for which the mean value of \vsini\ is 2 \kms, bias our results in the opposite sense and at least partially compensate for the other uncertainties. 

In our experiment, we used theoretical stellar evolution tracks to follow the individual evolution of planet hosting stars through the first ascent of the red giant branch.  We find that rapid rotators {\it can} be created from planet ingestion, though we find that the frequency of spin-up by planets can only account for $\sim$10\% of rapid rotator giants seen in the field.  
This disparity may be resolved if some of some of the RGB stars are in fact AGB or red clump stars. Even if our RGB stars are not confused, this disparity may simply indicate that other mechanisms, such as angular momentum dredge-up or low-mass stellar companion interactions, are also at work.
Contrary to our initial expectations, our modeling suggests that many of the known exoplanets will be ingested by their stars prior to the RGB, particularly those planets  in close-in orbits that have been found by transit searches.   Additionally, mass loss on the RGB as well as magnetic braking could significantly reduce the contribution of planet absorption to the formation of rapid rotators and thus require further scrutiny.
Nonetheless,  we find that the lower RGB is the best place to look for spin-up that is unambiguously due to planet accretion, which is possible both because the largest fraction of our predicted rapid rotators from planet accretion are rapidly rotating in this phase  and because first dredge-up, the time of possible angular momentum dredge-up from the core,  has not yet occurred.
Finally, there is growing evidence that relatively massive stars are more likely to harbor planets than less massive stars and that planets around these more massive stars have a different distribution of semimajor axes.  These trends  warrant further study, particularly since many red giant stars are more massive than the planet-hosting stars used in this experiment.

\vskip 0.1in
We thank the referee, Noam Soker, for a number of helpful suggestions that improved the presentation of our work. We gratefully acknowledge financial support by NASA Headquarters under the NASA Earth and Space Science Fellowship Program for project 08-Astro08F-0012 and by the F. H. Levinson Fund of the Peninsula Community Foundation. 


\begin{thebibliography}{38}
\expandafter\ifx\csname natexlab\endcsname\relax\def\natexlab#1{#1}\fi

\bibitem[{{Alexander}(1967)}]{alexander67}
{Alexander}, J.~B. 1967, Observatory, 87, 238

\bibitem[{{Carney} {et~al.}(2003){Carney}, {Latham}, {Stefanik}, {Laird}, \&
  {Morse}}]{carney03}
{Carney}, B.~W., {Latham}, D.~W., {Stefanik}, R.~P., {Laird}, J.~B., \&
  {Morse}, J.~A. 2003, \aj, 125, 293

\bibitem[{{Charbonnel} {et~al.}(1999){Charbonnel}, {D{\"a}ppen}, {Schaerer},
  {Bernasconi}, {Maeder}, {Meynet}, \& {Mowlavi}}]{charbonnel99}
{Charbonnel}, C., {D{\"a}ppen}, W., {Schaerer}, D., {Bernasconi}, P.~A.,
  {Maeder}, A., {Meynet}, G., \& {Mowlavi}, N. 1999, \aaps, 135, 405

\bibitem[{{de Medeiros} {et~al.}(1996){de Medeiros}, {Melo}, \&
  {Mayor}}]{deMed96a}
{de Medeiros}, J.~R., {Melo}, C.~H.~F., \& {Mayor}, M. 1996, \aap, 309, 465

\bibitem[{{Denissenkov} \& {Herwig}(2004)}]{denissenkov04}
{Denissenkov}, P.~A., \& {Herwig}, F. 2004, \apj, 612, 1081

\bibitem[{{Drake} {et~al.}(2002){Drake}, {de la Reza}, {da Silva}, \&
  {Lambert}}]{drake02}
{Drake}, N.~A., {de la Reza}, R., {da Silva}, L., \& {Lambert}, D.~L. 2002,
  \aj, 123, 2703

\bibitem[{{Fekel} \& {Balachandran}(1993)}]{fekel93}
{Fekel}, F.~C., \& {Balachandran}, S. 1993, \apj, 403, 708

\bibitem[{{Fischer} \& {Valenti}(2005)}]{fischer05}
{Fischer}, D.~A., \& {Valenti}, J. 2005, \apj, 622, 1102

\bibitem[{{Girardi} {et~al.}(2000){Girardi}, {Bressan}, {Bertelli}, \&
  {Chiosi}}]{giard00}
{Girardi}, L., {Bressan}, A., {Bertelli}, G., \& {Chiosi}, C. 2000, \aaps, 141,
  371

\bibitem[{{Glebocki} \& {Stawikowski}(2000)}]{glebocki00}
{Glebocki}, R., \& {Stawikowski}, A. 2000, Acta Astron., 50, 509

\bibitem[{{Gray}(1989)}]{Gray89}
{Gray}, D.~F. 1989, \apj, 347, 1021

\bibitem[{{Grether} \& {Lineweaver}(2006)}]{grether06}
{Grether}, D., \& {Lineweaver}, C.~H. 2006, \apj, 640, 1051

\bibitem[{{Jackson} {et~al.}(2008){Jackson}, {Greenberg}, \&
  {Barnes}}]{Jackson08}
{Jackson}, B., {Greenberg}, R., \& {Barnes}, R. 2008, \apj, 678, 1396

\bibitem[{{Johnson} {et~al.}(2007{\natexlab{a}}){Johnson}, {Butler}, {Marcy},
  {Fischer}, {Vogt}, {Wright}, \& {Peek}}]{Johnson07b}
{Johnson}, J.~A., {Butler}, R.~P., {Marcy}, G.~W., {Fischer}, D.~A., {Vogt},
  S.~S., {Wright}, J.~T., \& {Peek}, K.~M.~G. 2007{\natexlab{a}}, \apj, 670,
  833

\bibitem[{{Johnson} {et~al.}(2007{\natexlab{b}}){Johnson}, {Fischer}, {Marcy},
  {Wright}, {Driscoll}, {Butler}, {Hekker}, {Reffert}, \& {Vogt}}]{Johnson07}
{Johnson}, J.~A., et al. 2007{\natexlab{b}}, \apj, 665, 785

\bibitem[{{Johnson} {et~al.}(2006){Johnson}, {Marcy}, {Fischer}, {Henry},
  {Wright}, {Isaacson}, \& {McCarthy}}]{Johnson06}
{Johnson}, J.~A., {Marcy}, G.~W., {Fischer}, D.~A., {Henry}, G.~W., {Wright},
  J.~T., {Isaacson}, H., \& {McCarthy}, C. 2006, \apj, 652, 1724

\bibitem[{{Johnson} {et~al.}(2008){Johnson}, {Marcy}, {Fischer}, {Wright},
  {Reffert}, {Kregenow}, {Williams}, \& {Peek}}]{Johnson08}
{Johnson}, J.~A., {Marcy}, G.~W., {Fischer}, D.~A., {Wright}, J.~T., {Reffert},
  S., {Kregenow}, J.~M., {Williams}, P.~K.~G., \& {Peek}, K.~M.~G. 2008, \apj,
  675, 784

\bibitem[{{Livio} \& {Soker}(1984)}]{livio84}
{Livio}, M., \& {Soker}, N. 1984, \mnras, 208, 763

\bibitem[{{Livio} \& {Soker}(2002)}]{livio02}
---. 2002, \apjl, 571, L161

\bibitem[{{Massarotti}(2008)}]{massarotti08b}
{Massarotti}, A. 2008, \aj, 135, 2287

\bibitem[{{Massarotti} {et~al.}(2008){Massarotti}, {Latham}, {Stefanik}, \&
  {Fogel}}]{massarotti08a}
{Massarotti}, A., {Latham}, D.~W., {Stefanik}, R.~P., \& {Fogel}, J. 2008, \aj,
  135, 209

\bibitem[{{Murray} {et~al.}(2001){Murray}, {Chaboyer}, {Arras}, {Hansen}, \&
  {Noyes}}]{murray01}
{Murray}, N., {Chaboyer}, B., {Arras}, P., {Hansen}, B., \& {Noyes}, R.~W.
  2001, \apj, 555, 801

\bibitem[{{Pasquini} {et~al.}(2007){Pasquini}, {D{\"o}llinger}, {Weiss},
  {Girardi}, {Chavero}, {Hatzes}, {da Silva}, \& {Setiawan}}]{pasquini07}
{Pasquini}, L., {D{\"o}llinger}, M.~P., {Weiss}, A., {Girardi}, L., {Chavero},
  C., {Hatzes}, A.~P., {da Silva}, L., \& {Setiawan}, J. 2007, \aap, 473, 979

\bibitem[{{Peterson} {et~al.}(1983){Peterson}, {Tarbell}, \& {Carney}}]{pete83}
{Peterson}, R.~C., {Tarbell}, T.~D., \& {Carney}, B.~W. 1983, \apj, 265, 972

\bibitem[{{Reddy} {et~al.}(2002){Reddy}, {Lambert}, {Hrivnak}, \&
  {Bakker}}]{reddy02}
{Reddy}, B.~E., {Lambert}, D.~L., {Hrivnak}, B.~J., \& {Bakker}, E.~J. 2002,
  \aj, 123, 1993

\bibitem[{{Sahu} {et~al.}(2008){Sahu}, {Casertano}, {Valenti}, {Bond}, {Brown},
  {Smith}, {Clarkson}, {Minniti}, {Zoccali}, {Livio}, {Renzini}, {Rich},
  {Panagia}, {Lubow}, {Brown}, \& {Piskunov}}]{sahu07}
{Sahu}, K.~C., et al. 2008, in ASP Conf. Ser. 398, Extreme Solar Systems, ed. D.
Fischer, F. A. Rasio, S. E. Thorsett, and A. Wolszczan (San Francisco: ASP) 93

\bibitem[{{Sandquist} {et~al.}(1998){Sandquist}, {Taam}, {Lin}, \&
  {Burkert}}]{sandquist98}
{Sandquist}, E., {Taam}, R.~E., {Lin}, D.~N.~C., \& {Burkert}, A. 1998, \apjl,
  506, L65

\bibitem[{{Sandquist} {et~al.}(2002){Sandquist}, {Dokter}, {Lin}, \&
  {Mardling}}]{sandquist02}
{Sandquist}, E.~L., {Dokter}, J.~J., {Lin}, D.~N.~C., \& {Mardling}, R.~A.
  2002, \apj, 572, 1012

\bibitem[{{Siess} \& {Livio}(1999)}]{siess99}
{Siess}, L., \& {Livio}, M. 1999, \mnras, 308, 1133

\bibitem[{{Simon} \& {Drake}(1989)}]{simon89}
{Simon}, T., \& {Drake}, S.~A. 1989, \apj, 346, 303

\bibitem[{{Soker}(1996)}]{Soker96}
{Soker}, N. 1996, \apjl, 460, L53+

\bibitem[{{Soker}(1998)}]{soker98}
---. 1998, \aj, 116, 1308

\bibitem[{{Soker}(2004)}]{soker04}
{Soker}, N. 2004, in IAU Symposium, Vol. 219, Stars as Suns : Activity,
  Evolution and Planets, ed. A.~K. {Dupree} \& A.~O. {Benz} (Dordrecht: Kluwer), 323--+

\bibitem[{{Soker} \& {Harpaz}(2000)}]{soker00}
{Soker}, N., \& {Harpaz}, A. 2000, \mnras, 317, 861

\bibitem[{{Soker} \& {Harpaz}(2007)}]{soker07}
---. 2007, \apj, 660, 699

\bibitem[{{Sozzetti} {et~al.}(2009){Sozzetti}, {Torres}, {Latham}, {Stefanik},
  {Korzennik}, {Boss}, {Carney}, \& {Laird}}]{sozzetti09}
{Sozzetti}, A., {Torres}, G., {Latham}, D.~W., {Stefanik}, R.~P., {Korzennik},
  S.~G., {Boss}, A.~P., {Carney}, B.~W., \& {Laird}, J.~B. 2009, \apj, 697, 544

\bibitem[{{Verbunt} \& {Phinney}(1995)}]{vp95}
{Verbunt}, F., \& {Phinney}, E.~S. 1995, \aap, 296, 709

\bibitem[{{Zahn}(1977)}]{zahn77}
{Zahn}, J.-P. 1977, \aap, 57, 383

\end{thebibliography}



\end{document}